\documentclass[12pt,preprint]{aastex}

\usepackage{color}
\usepackage[breaklinks,colorlinks,citecolor=blue]{hyperref}

\shorttitle{A Binary Model for Repeating FRBs}
\shortauthors{Gu et al.}

\begin{document}

\title{A Neutron Star-White Dwarf Binary Model for \\
Repeating Fast Radio Burst 121102}

\author{Wei-Min Gu\altaffilmark{1}, Yi-Ze Dong\altaffilmark{1},
Tong Liu\altaffilmark{1,2}, Renyi Ma\altaffilmark{1},
and Junfeng Wang\altaffilmark{1}}

\altaffiltext{1}{Department of Astronomy, Xiamen University, Xiamen,
Fujian 361005, China; guwm@xmu.edu.cn}
\altaffiltext{2}{Department of Physics and Astronomy, University of Nevada,
Las Vegas, NV 89154, USA}

\begin{abstract}
We propose a compact binary model for the fast radio burst (FRB) repeaters,
where the system consists of a magnetic white dwarf (WD) and
a neutron star (NS) with strong bipolar magnetic fields.
When the WD fills its Roche lobe, mass transfer will occur
from the WD to the NS through the inner Lagrange point.
The accreted magnetized materials may trigger magnetic reconnection
when they approach the NS surface, and therefore the electrons can be
accelerated to an ultra-relativistic speed.
In this scenario, the curvature radiation of the electrons moving
along the NS magnetic field lines can account for the characteristic
frequency and the timescale of an FRB.
Owing to the conservation of angular momentum, the WD may be kicked away
after a burst, and the next burst may appear when the system becomes
semi-detached again through the gravitational radiation.
By comparing our analyses with the observations, we show that
such an intermittent Roche lobe overflow mechanism can be responsible
for the observed repeating behavior of FRB 121102.
\end{abstract}

\keywords{accretion, accretion disks --- binaries: general
--- gravitational waves --- magnetic reconnection --- white dwarfs}

\section{Introduction}

Fast radio bursts (FRBs) are millisecond-duration radio pulses with
mysterious physical origin.
With the exception of one FRB detected by the Arecibo telescope
\citep{Spitler2014} and one by the Green Bank Telescope (GBT), all the
observed FRBs were detected by the Parkes telescope
\citep[e.g.,][]{Lorimer2007,Keane2012,Thornton2013,Keane2016}.
The observed large values of dispersion measure indicate that FRBs
are probably extragalactic origin \citep[e.g.,][]{Katz2016}.
Several models have been introduced to interpret FRBs,
including giant flares from a magnetar \citep{Popov2010,Kulkarni2014},
giant pulses from a young pulsar or a magnetar
\citep{Cordes2016,Pen2015,Lyutikov2016},
the merger of neutron stars (NSs) or white dwarfs (WDs)
\citep{Totani2013,Kashiyama2013,WangJ2016},
the collapse of a massive NS into a black hole (BH)
\citep{Falcke2014,Zhang2014},
the collision between an NS and an asteroid or a comet \citep{Geng2015},
and the collapse of charged BHs \citep{Zhang2016a,Zhang2016b,Liu2016}.

Recently \citet{Spitler2016} showed ten additional bursts from FRB 121102
by Arecibo from May to June, 2015. Moreover, \citet{Scholz2016} showed six
additional bursts from November to December, 2015, among which five bursts
were detected by GBT and one by Arecibo. Thus, for this unique known
repeating FRB, there are totally 17 observed bursts as shown in Table~2 of
\citet{Scholz2016}. Obviously, the repeating bursts can rule out the models
involving cataclysmic events such as the merger of compact stars
or the collapse of a massive NS into a BH.
As mentioned in \citet{Scholz2016}, the probable extragalactic distance
and repeating behavior may favor giant pulses from a young pulsar
or a magnetar \citep{Cordes2016,Pen2015}, or radio counterparts to
magnetar X-ray bursts \citep{Lyutikov2002,Popov2013,Katz2015}.
In addition, \citet{Dai2016} suggested an interesting model that
a repeating FRB originates from a highly magnetized pulsar encountering
with lots of asteroids in an asteroid belt.
Very recently, \citet{WangF2016} studied the frequency distributions
of peak flux, fluence, duration and waiting time for this repeating source.

In this $Letter$, we propose a different model to understand the repeating
behavior of FRB 121102. Our binary system consists of a magnetized WD and an NS
with strong bipolar magnetic fields. We show that the intermittent Roche lobe
overflow mechanism can be responsible for the repeating behavior.
The remainder of this $Letter$ is organized as follows.
A compact binary model is illustrated in Section~2.
Analyses of the time interval between two adjacent bursts are studied
in Section~3.
Application of such a model to the repeating FRB 121102 is presented
in Section~4.
Conclusions and discussion are made in Section~5.

\section{Compact binary model}

In this section we propose a compact binary model for FRB repeaters.
The binary system consists of a magnetic WD and an NS with strong bipolar
magnetic fields. As illustrated in Figure~1, when the WD fills its
Roche lobe and the system is therefore called semi-detached, mass transfer
will occur from the WD to the NS. The materials in the atmosphere
(the yellow region) of the WD can pass through the inner Lagrange point
($L_1$ point) and then be accreted by the NS, as shown in Figure~1(a).
When the accreted magnetized
materials approach the NS surface, magnetic reconnection may be triggered
and therefore the electrons can be instantaneously accelerated to
an ultra-relativistic speed \citep[e.g.,][]{Zhang2011}.
Consequently, strong electromagnetic radiation can be released
by the curvature radiation as the electrons move along the NS
magnetic field lines.
On the other hand, as studied by \citet{King2007} based on the conservation
of angular momentum, the mass transfer will be a continuous behavior
for $q > 2/3$, where $q$ is defined as the mass ratio of the WD to the NS.
On the contrary, for $q < 2/3$, the WD may be kicked away
and the system may become detached after a Roche lobe overflow process,
as shown in Figure~1(b).
The system can become semi-detached again through the gravitational radiation
and therefore the next transfer process and corresponding
magnetic reconnection can restart.
Thus, for the case of $q < 2/3$, the Roche lobe overflow may be an
intermittent type.
Since the mass of an NS is normally larger than $1.4M_{\sun}$ and the 
mass distribution for WDs is peaked at $0.6M_{\sun}$, the intermittent
type of Roche lobe overflow may be a common phenomenon
for the semi-detached NS-WD systems.

We would point out that the cataclysmic variable (CV) is an analogous system.
However, accretion in CVs is continuous rather than
episodic. In our opinion, the main difference is that, in our model
the accreted materials are from a WD instead of a main sequence star.
We may therefore expect a much more violent mass transfer than in CVs.
As a consequence, the outward moving due to the conservation of angular
momentum may dominate over the inward moving owing to the gravitational
radiation, and therefore the WD may be kicked away after a mass transfer
process. We would stress that the violent mass transfer and the corresponding
kick-away behavior is a fundamental assumption in our model.

The characteristic frequency of the curvature radiation
of relativistic electrons is expressed as
\begin{equation}
\nu_{\rm c} = \frac{3c \gamma^3}{4\pi R_{\rm c}}
= 1.5 \left( \frac{\gamma}{60} \right)^3
\left( \frac{R_{\rm c}}{10^6 {\rm cm}} \right)^{-1}~{\rm GHz} \ ,
\end{equation}
where $\gamma$ is the Lorentz factor of electrons, and $R_{\rm c}$ is
the curvature radius. The above equation means that, for a typical radius
$R_{\rm c} = 10^6 {\rm cm}$, an FRB with $\nu_{\rm c}$ of order
$\sim$ GHz requires $50 \la \gamma \la 100$, which is a reasonable
range according to the acceleration by magnetic reconnection
\citep[e.g.,][]{Zhang2011,Kowal2012}.

The duration of an FRB may be regarded as around the timescale
of a magnetic reconnection.
According to our model, the energy-released region is near the NS surface,
and the duration $t_{\rm w}$
can be estimated by the ratio of
the NS radius $R_{\rm NS}$ to the Alfv\'{e}n speed
$v_{\rm A}(=B_{\rm NS}/\sqrt{4\pi \bar\rho})$:
\begin{equation}
t_{\rm w} = \frac{R_{\rm NS}}{v_{\rm A}}
= 1.1 \left( \frac{R_{\rm NS}}{10^6 {\rm cm}} \right)
\left( \frac{B_{\rm NS}}{10^{11} {\rm G}} \right)^{-1}
\left( \frac{\bar\rho}{10^3 {\rm g~cm^{-3}}} \right)^{\frac{1}{2}}~{\rm ms} \ ,
\end{equation}
where $B_{\rm NS}$ is the magnetic flux density, and $\bar\rho$ is the
averaged mass density of accreted materials.
Equation~(2) shows that, for typical values
$R_{\rm NS}=10^6 {\rm cm}$, $B_{\rm NS} = 10^{11} {\rm G}$, and
$\bar\rho = 10^3 {\rm g~cm^{-3}}$, an FRB will be a millisecond duration.

Here, we would stress that the mass density in the atmosphere of a WD varies
continuously from large interior values to essentially zero.
In our analyses, for simplicity, we adopt an averaged mass density
$\bar\rho = 10^3 {\rm g~cm^{-3}}$ for the accreted materials.
In Section~5, we will discuss the degeneracy of $\bar\rho$ with other
physical quantities in our model.

\section{Time interval analyses}

In this section we investigate the analytic variation of the time interval 
between two bursts with the transferred mass during the former burst.
We assume a circular orbit for the analyses, i.e., the eccentricity $e=0$.
The orbital angular momentum $J$ of a binary system takes the form
\begin{equation}
J = M_1 M_2 \left( \frac{Ga}{M} \right)^{\frac{1}{2}} \ ,
\end{equation}
where $M_1$ and $M_2$ are respectively the NS and WD mass,
$M = M_1 + M_2$ is the total mass, and $a$ is the binary separation. 
When the WD fills its Roche lobe, mass transfer will occur from the WD
to the NS. We assume $\Delta M_2$ as the transferred mass during one
burst, where $\Delta M_2$ is negative. Then, the orbital angular
momentum carried by the accreted materials can be expressed as
\begin{equation}
\Delta J = -\lambda \Delta M_2 \Omega (b_1 - a_1)^2 \ ,
\end{equation}
where $\Omega$ is the orbital angular velocity, $a_1$ is the distance
between the NS and the center of the mass, $b_1$ is the distance
between the NS and the $L_1$ point, and $\lambda$ is a parameter
probably in the range $0 \leqslant \lambda \leqslant 1$.
The expression of $b_1$ takes the form \citep[e.g., Chap.~4.4 of][]{Frank2002}
\begin{equation}
\frac{b_1}{a} = 0.5 - 0.227\log{q} \ ,
\end{equation}
where $q \equiv M_2/M_1$.
For $0.1 \la q \la 0.8$ it is often convenient to adopt a simple form
to describe the Roche lobe radius of the secondary $M_2$:
\begin{equation}
\frac{R_2}{a} = 0.462 \left( \frac{M_2}{M} \right)^{\frac{1}{3}} \ .
\end{equation}
The radius of a WD is expressed as
\begin{equation}
R_{\rm WD} = 10^9 \left( \frac{1.402\times 10^{33}~{\rm g}}{M_2} \right)
^{\frac{1}{3}} \ {\rm cm} \ .
\end{equation}
In our scenario, the mass transfer will occur for $R_{\rm WD} \ga R_2$.
After a mass transfer process, the WD may be kicked away and the
next transfer can appear through the gravitational radiation.
For the simple case with $e=0$, the variation of $a$ due to the
gravitational radiation can be written as \citep{Peters1964}
\begin{equation}
\frac{da}{dt} = -\frac{64}{5} \frac{G^3 M_1 M_2 M}{c^5 a^3} \ .
\end{equation}
Based on Equations (3-8), we can derive an analytic relation between
the time interval $\Delta t$ between two adjacent bursts and
the transferred mass $\Delta M_2$ during the former burst,
\begin{equation}
\Delta t = \frac{5 (1+q) c^5 a^4}{32 q G^3 M^3}
\left\{ \lambda \left[ 0.5(1-q) - 0.227(1+q) \log{q} \right]^2
- (1+q)(\frac{2}{3} - q) \right\} \frac{\Delta M_2}{M_2} \ .
\end{equation}
The above equation enables us to obtain $\Delta t$ once $\Delta M_2$ is given.

\section{Application to FRB 121102}

As mentioned in Section~1, for the unique known repeating FRB 121102,
there are totally 17 observed bursts, as shown in Table~2 of
\citet{Scholz2016}.
In this section, we focus on the time intervals between two adjacent
bursts measured during periods of continuous observation.
Based on the data from \citet{Spitler2016} and \citet{Scholz2016},
we can derive totally ten such intervals among the 17 bursts, as shown
by our Table~1. It is seen that the ten intervals are from tens to hundreds
of seconds. According to our scenario, an FRB is related to a magnetic
reconnection event.
The physical mechanism of such reconnection may be
attributed to a self-organized criticality process \citep{Wang2013}.

Now we can compare our analyses with the observations of FRB 121102,
mainly on the relation between the time interval $\Delta t$ and
the transferred mass $\Delta M_2$.
In our model we adopt typical masses for the NS and the WD, i.e.,
$M_1 = 1.4 M_{\sun}$ and $M_2 = 0.6 M_{\sun}$.
The corresponding separation for the critical situation
$R_{\rm WD} = R_2$ is $a = 3.41\times 10^9$~cm.
There are two extreme cases that the accreted materials do not carry
the orbital angular momentum ($\lambda = 0$), and the materials
carry the Keplerian angular momentum at the $L_1$ point ($\lambda = 1$).
Previous studies on CVs indicate that $\lambda$ is 
around unity. We therefore adopt $\lambda = 1$ in the following analyses.
Then, Equation~(9) can be simplified as
\begin{equation}
\Delta t = - 2.68 \times 10^{10} \frac{\Delta M_2}{M_{\sun}} \ {\rm s} \ .
\end{equation}

We try to show a comparison between the analyses and
the observations based on two observational quantities, i.e.,
the fluence $F$ and the time interval $\Delta t$.
A strong magnetic field with $B_{\rm WD} \approx 3\times 10^8$G is required
for the WD (magnetic fields have been discovered in over 100 WDs,
ranging from $10^3$G to $10^9$G).
The averaged mass density of the atmosphere is adopted as
$\bar\rho = 10^3$g~cm$^{-3}$.
Here, we choose $z=0.1$ according to the DM measurement,
and the luminosity distance is $D_{\rm L} = 463.4~{\rm Mpc}$
for $H_0=69.6$, $\Omega_{\rm M}=0.286$, and $\Omega_{\rm vac}=0.714$.
In our model, it is difficult to estimate how many electrons have been
accelerated to an ultra-relativistic speed. Alternatively, we try to evaluate
the released energy by the magnetic energy carried by the accreted materials.
In a real case, the magnetic diffusion and the magnetorotational instability
(MRI) may occur in the accretion flow and therefore change the magnetic energy.
Here, for simplicity, we just assume that the carried magnetic energy
keeps unchanged during the accretion process.
Thus, we can evaluate the fluence $F$ by the total magnetic
energy carried by the accreted materials:
\begin{equation}
F \Delta\nu D_{\rm L}^2 \Delta\theta
= - \eta \frac{B_{\rm WD}^2}{8\pi} \frac{\Delta M_2}{\bar\rho} \ ,
\end{equation}
where $\Delta\nu$ is the width of radio frequency,
$\Delta\theta$ is the solid angle for an FRB,
and $\eta$ is the efficiency of the released energy through
the curvature radiation compared with the total magnetic energy
carried by the accreted materials.
We adopt $\Delta\nu = 1$GHz and $\Delta\theta = 0.04\pi$
(corresponding to 1\% of the whole space) for the calculations.

A comparison of analyses with observations in the $\Delta t-F$ diagram
is shown in Figure~2, where $\Delta t$ is the time interval in the rest frame.
The three solid lines represent the analytic
relation for the efficiency $\eta = 0.001$, 0.01, and 0.1,
which are derived by combining Equations (10-11).
The ten color filled circles denote the
observational results shown in Table~1.
It is seen from Figure~2 that all the bursts are well
located among the three theoretical lines, which implies that the analyses
are roughly in agreement with the observations.
Thus, the NS-WD binary model is likely to be responsible for the FRB repeaters.

\section{Conclusions and discussion}

In this $Letter$ we have proposed an NS-WD binary model for the FRB repeaters.
The system consists of a magnetic WD and an NS with strong bipolar
magnetic fields.
Mass transfer will occur through the $L_1$ point
when the WD fills its Roche lobe.
Consequently, the accreted magnetized materials may trigger
magnetic reconnection when they approach the NS surface, and therefore
the electrons can be accelerated to an ultra-relativistic speed.
In this scenario, an FRB can be powered by the curvature radiation
of the relativistic electrons moving along the NS magnetic field lines.
Owing to the conservation of angular momentum, the WD may be
kicked away after a burst, and the next burst may appear when
the system becomes semi-detached again through the gravitational radiation.
Our analyses have shown that such an intermittent Roche lobe overflow
mechanism can be responsible for the repeating behavior of FRB 121102.

For the application of our model to FRB 121102, we would point out that
the required values for physical quantities are degenerate, as indicated
by Equations (2) and (11). For instance, if the averaged mass density is
$\bar\rho = 10 {\rm g~cm^{-3}}$ instead of $10^3 {\rm g~cm^{-3}}$,
these two equations can be again satisfied with $B_{\rm NS} = 10^{10}$G
and $B_{\rm WD} = 3\times 10^7$G. In other words, our model can work
for the repeating FRB with another group of parameters. Actually,
as shown by Equation~(11), either a smaller $\eta$ or a larger
distance $D_{\rm L}$ may be equivalent to a smaller $B_{\rm WD}$.
Thus, Section~4 just shows an example group of parameters
which can work for FRB 121102.
In addition, we would again stress that a fundamental assumption
in our model is that, once the Roche lobe is filled, the mass transfer will
be so violent that the WD can be kicked away after a mass transfer process.

In this work we have focused on the curvature radiation which
mainly contribute to the radio emission. However,
the synchrotron radiation may also be of importance for relativistic
electrons moving in magnetic fields. In the case of $B_{\rm NS}=10^{11}$G
for the NS,
the characteristic frequency for the synchrotron radiation is
$\nu_{\rm s} = 3\gamma^3 eB_{\rm NS}/4\pi m_{\rm e}c \sim 1{\rm MeV}$ for
$\gamma \sim 60$.
Since the energy of curvature (synchrotron) radiation is related to
the component of kinetic energy of electrons parallel (perpendicular)
to the magnetic field lines, we can regard the energy budget for
these two radiation mechanisms is comparable.
As a consequence, it is easy to estimate that
the fluence of 1Jy~ms with 1GHz width is equivalent to $\sim 0.06$eV
for the total energy with a square meter high-energy detector, which means
that the probability of receiving one MeV photon by such a detector
is less than $10^{-7}$.
Thus, the possible gamma-ray emission by the synchrotron radiation
can hardly be detected except that the FRB occurs in the Local Group.

As mentioned in Section~1, \citet{Dai2016} introduced a model related
to asteroids. Their radiation mechanism is also the curvature radiation.
The difference is that, the energy source for FRBs in their model
is the kinetic energy of asteroids, whereas the source in our model
is the magnetic energy of accreted materials. In our scenario, it is
still possible for a part of the kinetic energy of accreted materials to
be transferred to the electrons through the strong magnetic fields,
and therefore have contribution to the radio emission.
In such case, the required magnetic fields of the WD
may be significantly weaker than the present assumption.

\acknowledgments

We thank Zhongxiang Wang, Bing Zhang, and Zi-Gao Dai for beneficial
discussions, and the referee for constructive suggestions that
improved this $Letter$.
This work was supported by the National Basic Research Program of China
(973 Program) under grant 2014CB845800,
the National Natural Science Foundation of China under grants 11573023,
11522323, 11473022, 11473021, 11333004, 11222328, U1531130, and U1331101,
and the Fundamental Research Funds for the Central Universities
under grants 20720140532, 20720150024, 20720160023, and 20720160024.

\clearpage

\begin{figure}
\plotone{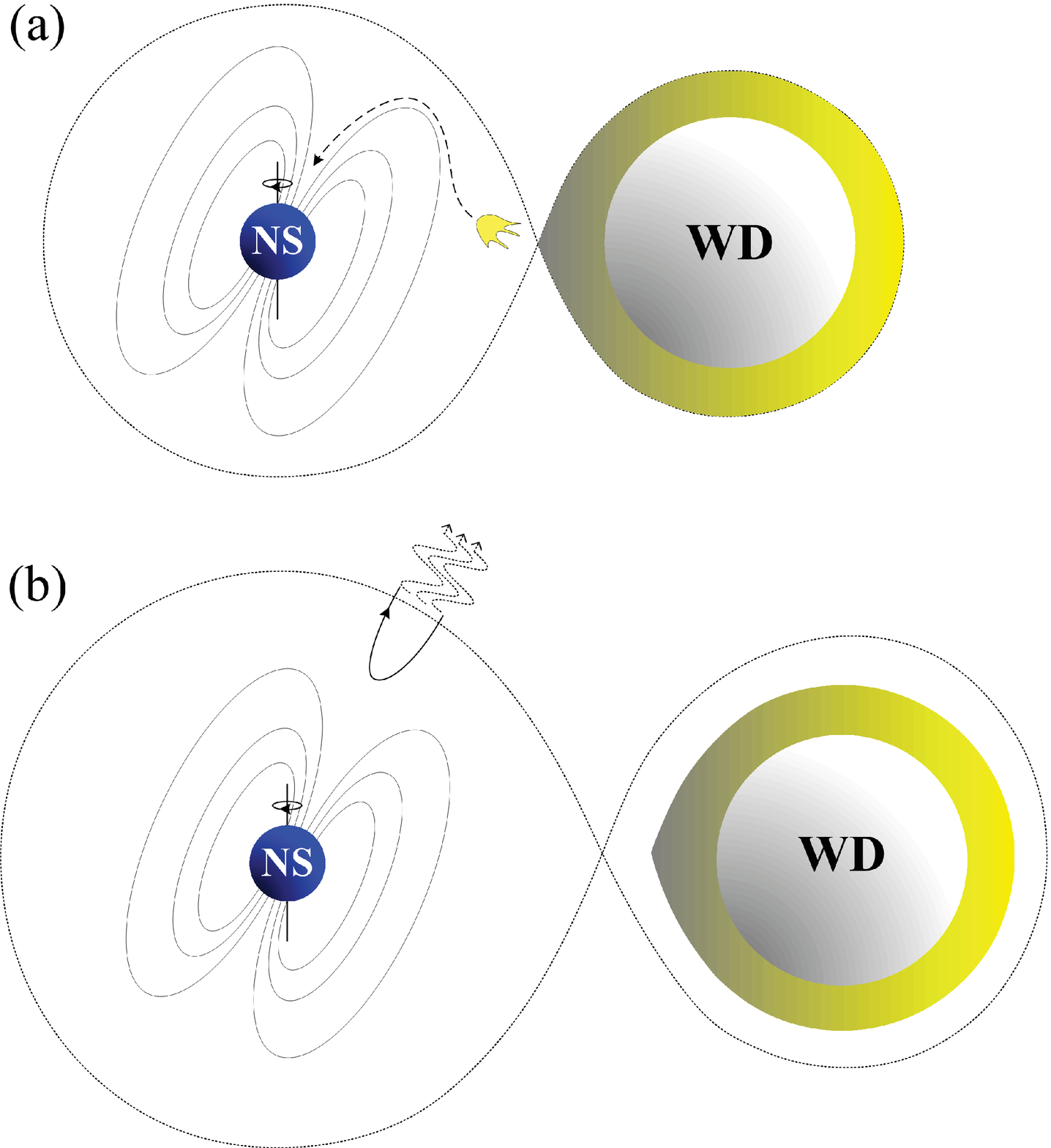}
\caption{
Illustration of the intermittent Roche lobe overflow in an NS-WD binary system:
(a) the WD fills its Roche lobe and mass transfer occurs through the
inner Lagrange point; (b) the WD is kicked away after the mass transfer,
and the accreted materials trigger magnetic reconnection and strong
electromagnetic radiation.
}
\end{figure}

\clearpage

\begin{figure}
\plotone{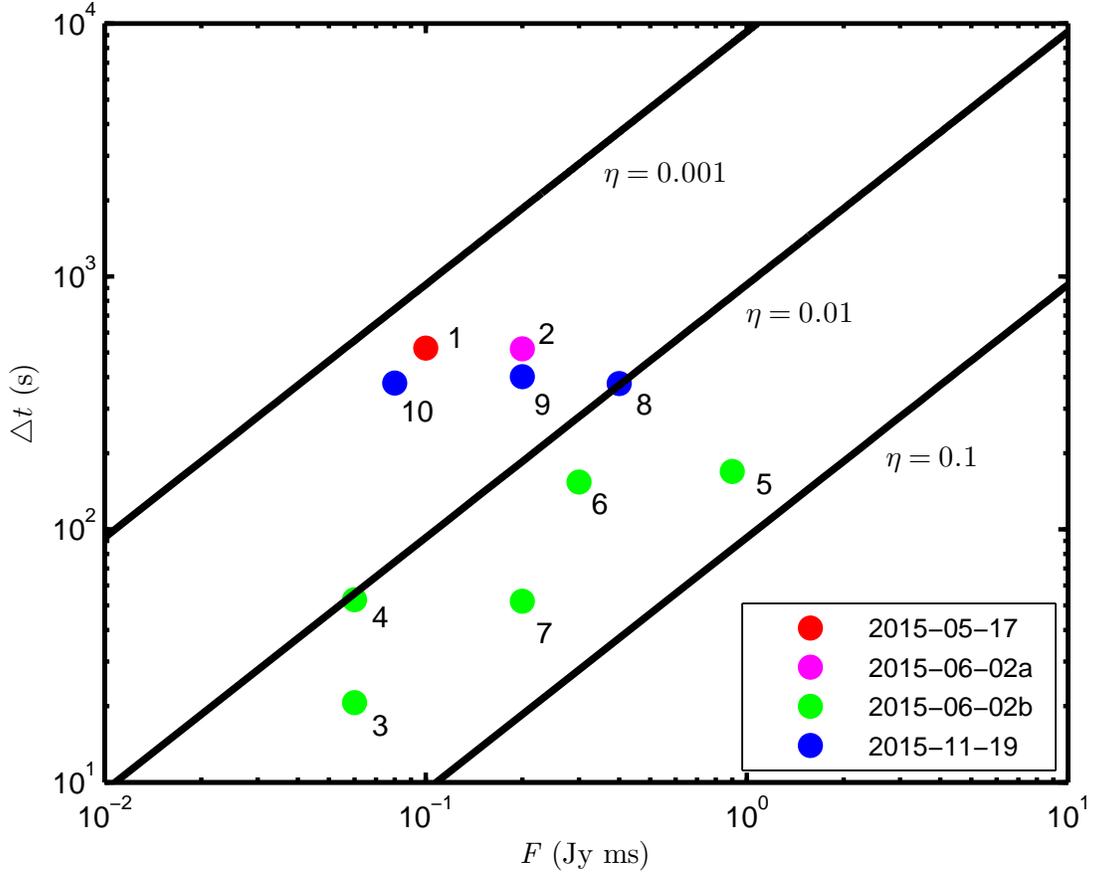}
\caption{
Comparison of analyses with observations in the $\Delta t - F$ diagram.
The three solid lines represent the analytic relation for $\eta = 0.001$,
0.01, and 0.1,
and the other parameters in Equation~(11) are fixed as $\Delta\nu = 1$GHz,
$D_{\rm L} = 463.4~{\rm Mpc}$ ($z=0.1$), $\Delta\theta = 0.04\pi$,
$B_{\rm WD} = 3\times 10^8$G, and $\bar\rho = 10^3$g~cm$^{-3}$.
The ten color filled circles denote the observational results shown in Table~1.
}
\end{figure}

\clearpage

\begin{deluxetable}{clcc}
\tabletypesize{\normalsize}
\tablecaption{Ten intervals in continuous observation of FRB 121102}
\tablewidth{373pt}
\tablehead{\colhead{Interval number} & \colhead{Date} &
\colhead{Fluence (Jy ms)} & \colhead{Time interval (s)} }
\startdata
1 & 2015-05-17 & 0.1 & 572.2 \\
2 & 2015-06-02a & 0.2 & 568.9 \\
3 & 2015-06-02b & 0.06 & 22.7 \\
4 & 2015-06-02b & 0.06 & 58.0 \\
5 & 2015-06-02b & 0.9 & 186.2 \\
6 & 2015-06-02b & 0.3 & 169.3 \\
7 & 2015-06-02b & 0.2 & 57.2 \\
8 & 2015-11-19 & 0.4 & 414.4 \\
9 & 2015-11-19 & 0.2 & 441.3 \\
10 & 2015-11-19 & 0.08 & 416.3
\enddata
\tablenotetext{{\rm Note}}{Adopted from \citet{Spitler2016} and
\citet{Scholz2016}.}
\label{table1}
\end{deluxetable}

\end{document}